\documentclass[conference]{IEEEtran}
\IEEEoverridecommandlockouts
\usepackage{cite}
\usepackage{amsmath,amssymb,amsfonts}
\usepackage{algorithmic}
\usepackage{graphicx}
\usepackage{textcomp}
\usepackage{xcolor}
\usepackage[font=normalsize]{caption}
\usepackage{caption}
\usepackage{subcaption}
\usepackage{amsthm}

\newtheorem{invariant}{Invariant}
\usepackage{array}
\usepackage{url}

\definecolor{bluekeywords}{rgb}{0,0,1}
\definecolor{greencomments}{rgb}{0,0.5,0}
\definecolor{redstrings}{rgb}{0.64,0.08,0.08}
\definecolor{xmlcomments}{rgb}{0.5,0.5,0.5}
\definecolor{types}{rgb}{0.17,0.57,0.68}

\usepackage{listings}
\renewcommand{\baselinestretch}{0.922}
\newcommand{\TODO}[1]{\textsf{\textbf{[TODO: }\textsl{#1}\textbf{]}}}
\def\BibTeX{{\rm B\kern-.05em{\sc i\kern-.025em b}\kern-.08em
    T\kern-.1667em\lower.7ex\hbox{E}\kern-.125emX}}
\begin{document}

\newcommand{\R}{\mathbb{R}}
\newcommand{\K}{P}
\newcommand{\I}{U}
\newcommand{\C}{C}
\newcommand{\A}{\mathcal{A}}
\newcommand{\U}{\mathcal{U}}
\newcommand{\Po}{\mathcal{P}}
\newcommand{\T}{\mathcal{T}}

\newcommand{\E}{\mathbb{E}}
\renewcommand{\L}{\mathcal{L}}
\newcommand{\Lq}{\mathcal{L}}
\title{{\sc BakUp}: Automated, Flexible, and Capital-Efficient Insurance Protocol for Decentralized Finance
}
\newcommand{\name}{\sc BakUp}

\author{\IEEEauthorblockN{Srisht Fateh Singh}
\IEEEauthorblockA{\textit{Electrical and Computer Engineering} \\
\textit{University of Toronto}\\
Toronto, Canada \\
srishtfateh.singh@mail.utoronto.ca}
\and
\IEEEauthorblockN{Panagiotis Michalopoulos}
\IEEEauthorblockA{\textit{Electrical and Computer Engineering} \\
\textit{University of Toronto}\\
Toronto, Canada \\
p.michalopoulos@mail.utoronto.ca}
\and
\IEEEauthorblockN{Andreas Veneris}
\IEEEauthorblockA{\textit{Electrical and Computer Engineering} \\
\textit{University of Toronto}\\
Toronto, Canada \\
veneris@eecg.toronto.edu}
}
\renewcommand{\baselinestretch}{0.95}
\maketitle

\begin{abstract}
This paper introduces {\name}, a smart contract insurance design for decentralized finance users to mitigate risks arising from platform vulnerabilities. While providing automated claim payout, {\name} utilizes a modular structure to harmonize three key features: the platform's resilience against vulnerabilities, the flexibility of underwritten policies, and capital efficiency. An immutable core module performs capital accounting while ensuring robustness against external vulnerabilities, a customizable oracle module enables the underwriting of novel policies, and an optional and peripheral yield module allows users to independently manage additional yield. The implementation incorporates binary conditional tokens that are tradable on automated market maker (AMM)-based exchanges. Finally, the paper examines specific liquidity provision strategies for the conditional tokens, demonstrating that a conservative strategy and parameterization can effectively reduce the divergence loss of liquidity providers by more than $47\%$ compared to a naive strategy in the worst-case scenario.
\end{abstract}

\section{Introduction}
Programmable blockchains~\cite{nakamoto2008peer,buterin2014ethereum, avalanche,yakovenko2018solana, nightshade} have led to an era of decentralized finance, or DeFi, where centralized intermediaries are replaced by software code, 
also known as \emph{smart contracts}, to provide  
financial services. This has spawned novel financial innovations such as automated execution, resulting in no market downtime, and permissionless access, increasing user inclusiveness and applications' interoperability compared to its traditional (TradFi) counterpart. This paradigm shift has led to increased utilization of DeFi systems in recent years, with several billion dollars exchanging hands daily~\cite{defiVolume}. 

Despite its success, DeFi is not free from risks, thus impeding its rapid adoption~\cite{atzei2017survey, gudgeon2020decentralized,zhou2023sok}. These risks can generally be classified as {\em technical} or {\em economic}~\cite{werner2022sok}. Technical risks arise from exploiting the vulnerabilities of
smart contracts, such
as programming bugs and/or unintentional functional specification inconsistencies that can lead to irreversible loss of funds held by a DeFi protocol. 
The second type of risks,  stem from economic inefficiencies within a protocol. Technical and economic exploitation, along with external factors, often lead to unexpected behavior in DeFi metrics. This includes fluctuations in token prices and liquidity in automated market-making (AMM) pools, as well as extreme borrowing rates in lending protocols\footnote{https://finance.yahoo.com/news/defi-lenders-spooked-curve-exploit-193953614.html}. Such deviations can rupture the functioning of dependent protocols~\cite{bekemeier2021deceptive} or result in losses for end users. For example, the liquid staking protocol Lido Finance~\cite{lido2020} has a wrapped token for staked ETH (stETH), which is expected to maintain its price stability relative to ETH. However, if this characteristic fails due to the aforementioned reasons, stETH investors may incur unforeseen losses. This emphasizes the imperative need for robust {\em DeFi risk management tools} to safeguard users from such consequential risks.

One category of solutions to the above problem involves insurance, where the risk of users (policyholders) is transferred to underwriters. However, the
very nature of DeFi demands an insurance platform that is (a) automated, ensuring instantaneous claim payout; (b) permissionless in access to achieve interoperability; (c) resilient against both technical and economic vulnerabilities, as it is the last resort for financial adversities; (d) flexible to underwrite a vast pool of policies; and (e) capital-efficient, \emph{i.e.}, any locked capital should be capable of earning yield.


An insurance platform completely implemented using smart contracts achieves the first two properties, \emph{viz.,} automation and permissionless access. However, the latter three merits, \emph{viz.,} resilience, flexibility, and capital-efficiency {\em inherently} compromise the objectives of each other. For instance, a high level of resilience in contract design implies high immutability, which in turn hinders the underwriting flexibility derived from adaptability or upgradability. Simultaneously, the resilience of the smart contract platform may not align well with capital efficiency, as yield generation often involves interactions with third-party protocols that are susceptible to vulnerabilities. 

This paper proposes {\name}, a design for a modular, flexible, and capital-efficient smart contract platform that uses binary conditional tokens to provide insurance policies. {\name} allows users to independently align between resilience, flexibility, and capital efficiency by comprising three distinct modules: a {\em core module} for basic capital accounting to ensure no defaults, an {\em oracle module} for underwriting policies, and a {\em yield module } on the periphery of the above two for capital management. The core module is immutable, while the oracle module can be created permissionlessly with a custom developer-defined logic, which can range from conservative to highly customizable. Simultaneously, the peripheral yield module operates independently from the main protocol, making it optional for users to engage in yield-generation activities. Interestingly, this optional feature doesn't pose any risk to the capital of users who choose not to participate, allowing users—both underwriters and policyholders—with varying risk preferences to interact simultaneously on the platform without imposing risks on others.

Here, we first present the design of the core module. Subsequently, we describe the oracle module and demonstrate a contract design capable of underwriting the de-pegging of stable assets using a combination of on-chain and off-chain price sources. Thereafter, we present the yield module's design that integrates the DeFi Risk Transfer protocol~\cite{nadler2023defi} for secure yield generation, allowing users to mitigate their technical risks by paying an additional premium. This allows users to choose between unsecured yield (high-risk/reward), secure yield (medium-risk/reward), or no yield at all (low-risk/reward). 
The insurance policies, implemented as a binary conditional ERC20 token, can be traded on both decentralized (DEX) and centralized exchanges (CEX). As such, this paper also presents a DEX marketplace by analyzing the divergence loss of liquidity providers in three broad strategies for the passive liquidity provision of these tokens. Empirical results show that conservative strategies and parameterization can help mitigate the loss of liquidity providers by over $47\%$ compared to the naive strategy in the worst-case scenario.

This paper is organized as follows. Section~\ref{sec:background} gives background knowledge, Section~\ref{sec:design} presents the design and implementation of the various modules, Section~\ref{sec:marketplace} gives a market analysis of the underlying tokens, including liquidity provision strategies, that
are later evaluated in Section~\ref{sec:evaluation}. Section~\ref{sec:related} discusses related works, and Section~\ref{sec:conclusion} concludes this work.

\section{Background}\label{sec:background}
\subsection{DeFi Risk Transfer}\label{sec:DRT}
Yield-generating protocols such as Aave~\cite{aave} and Compound~\cite{leshner2019compound} are susceptible to risks arising from platform vulnerabilities. Thus, investing in these platforms carries a high probability of net positive yield and a low probability of net negative yield, the latter being attributed to platform vulnerabilities. DeFi Risk Transfer (DRT)~\cite{nadler2023defi} is a protocol inspired by collateral debt obligations (CDOs)~\cite{duffie2001risk} that allows investors interested in earning yield on an underlying asset $\C$, such as USDC, to distribute the risk between risk-averse and risk-taking investors. The former can transfer the risk of these low-probability adverse events to the latter by paying a premium. DRT accomplishes this through the following steps.
\begin{enumerate}
    \item \emph{Tranche Creation:} DRT creates two tranches: a low-risk $A$-tranche with members holding an ERC20 token $A$ and a high-risk $B$-tranche with members holding token $B$. A user depositing $1$ unit of $\C$ mints $1$ unit each of $A$ and  $B$.
    
    \item \emph{Investing:} Given $k$ units of $\C$ as the total deposited capital, DRT equally invests this ($k/2$ units each) between two yield-generating platforms. In return, it receives yield-bearing tokens $\C_{1}$ and $\C_{2}$ respectively, serving as receipts. This capital is kept invested until a set deadline.
    \item \emph{Risk Splitting and Divesting:} After the deadline passes, DRT divests the capital by exchanging $\C_{1}$ and $\C_{2}$ at the first and second platform, respectively. Denoting $k'$ as the total units of $\C$ redeemed and $k_{i}$ as the interest accrued for each tranche, the capital is split between the two tranches based on one of the three scenarios: 
    \begin{itemize}
        \item \textbf{$k' \ge k:$} In the normal scenario, where DRT receives its entire capital plus additional yield, both $A$- and $B$-tranche receives $k'/2$ units each. This capital is then uniformly distributed among holders of $A$ and $B$.
        \item \textbf{$k > k'> k/2:$} In an adverse scenario where DRT receives only a partial fraction of the invested amount, $A$-tranche
        receives $k/2 + k_{i}$ units, 
        while $B$-tranche receives $k' - (k/2 + k_{i})$ units of $\C$. Therefore, the risk is transferred from $A$- to $B$-tranche.
        \item \textbf{$k' \le k/2:$} In an extremely adverse scenario where DRT receives less than half of its initial investment, the entire capital, \emph{i.e.,} $k'$ units, is received solely by
        $A$-tranche.
    \end{itemize}
    \item \emph{Pricing $A$ and $B$:} External markets for $A$ and $B$ are established by creating a trading pair $A/B$ on AMM-based DEXs. Since the expected pay-off of $A$ is higher than $B$, the fair price of $A$ in terms of $B$ is greater than $1$ and is decided by the market. 
    \item \emph{End user:} A user willing to acquire either only $A$ or $B$ can first mint equal amounts of $A$ and $B$ and then exchange the other token for the desired token. For example, a user willing to invest $1$ unit of $\C$ in $A$-tranche first mints $1$ unit of $A$ and $B$ and then swap $B$ in exchange for $A$ at the AMM marketplace.
\end{enumerate}
Since the architecture of DRT only transfers the risk related to yield-generating platforms, it misses out on the flexibility of underwritten policies.
\subsection{Concentrated Liquidity Automated Market Makers}\label{sec:conc-liq}
In the context of Concentrated Liquidity Automated Market Makers (CLAMM)~\cite{adams2021uniswap}, we consider a pair of tokens $X$ and $Y$ and measure the price of token $X$ in terms of $Y$. When providing liquidity for this pair, a liquidity provider (LP) chooses a price  interval $[p_{0},p_{1}]$ with utility as
explained in the next paragraph. This, along with the current price $p$, determines the ratio of the two tokens that the LP needs to deposit. In turn, the amount of tokens that the LP actually deposits (in accordance with the above ratio) determines the liquidity $l$ provisioned to the interval.

The current price of the pool changes with  each trade. The corresponding trading fees are distributed to all LPs in proportion to their liquidity in the interval containing the current price. If the price moves outside the chosen interval of an LP, their position consists of only one of the two tokens, and they stop earning fees. However, this resumes if the price returns to within their interval.
Given a fixed amount of tokens that the LP wants to provision, the size of the price interval affects the proportion of fee earned. Depositing the tokens in a narrower interval increases LPs proportion of fees but leads to greater LP losses when price changes~\cite{loesch2021impermanent}.

Considering a liquidity position in the interval $[p_{0},p_{1}]$, let $r'_{X}(p)$ and $r'_{Y}(p)$ denote the \emph{virtual reserves} of the position at a price of $p\in[p_{0},p_{1}]$, \emph{i.e.,} the amounts of tokens used to compute an incoming trade. Then $r'_{X}(p)r'_{Y}(p)=l^2$ and $r'_{Y}(p)/r'_{X}(p)=p$. This implies $r'_{X}(p)=l/\sqrt{p}$ and $r'_{Y}=l\sqrt{p}$. Therefore, the amount of tokens that the position actually consists of or the \emph{real reserves} $r_{X}(p)$ and $r_{Y}(p)$ can be calculated as follows:
\begin{equation}
\begin{split}
    r_{X}(p) = r'_{X}(p)- r'_{X}(p_1)=l\cdot(\frac{1}{\sqrt{p}}-\frac{1}{\sqrt{p_{1}}})\\
    r_{Y}(p) = r_{Y}(p)-r'_{Y}(p_0) = l\cdot(\sqrt{p}-\sqrt{p_{0}})
\end{split}
\end{equation}
Note that the above equations only hold for $p\in[p_0,p_1]$. Otherwise, we have $r_{X}(p)=r_{X}(p_{0})$ and $r_{Y}(p)=0$ for $p<p_{0}$, and $r_{X}(p)=0$ and $r_{Y}(p)=r_{Y}(p_{1})$ for $p>p_{1}$. Therefore, an LP entering a position in the above range with $p<p_{0}$ and exiting with $p>p_{1}$ is analogous to swapping $r_{X}(p_{0})$ of $X$ for $r_{Y}(p_{1})$ of $Y$. The average execution price of this trade is therefore:
\begin{equation}
  \frac{r_{Y}(p_{1})}{r_{X}(p_{0})} =  \frac{l\cdot(\sqrt{p_{1}}-\sqrt{p_{0}})}{l\cdot(\frac{1}{\sqrt{p_{0}}}-\frac{1}{\sqrt{p_{1}}})} = \sqrt{p_{0}p_{1}}
\end{equation}
which is the geometric mean of the extreme price of the LP's position. We will use this result to evaluate liquidity provision strategies in Section~\ref{sec:evaluation}.

\section{\name \ Protocol Design}\label{sec:design}

\subsection{Protocol Overview}
\begin{figure}[t]
\includegraphics[width=9cm]{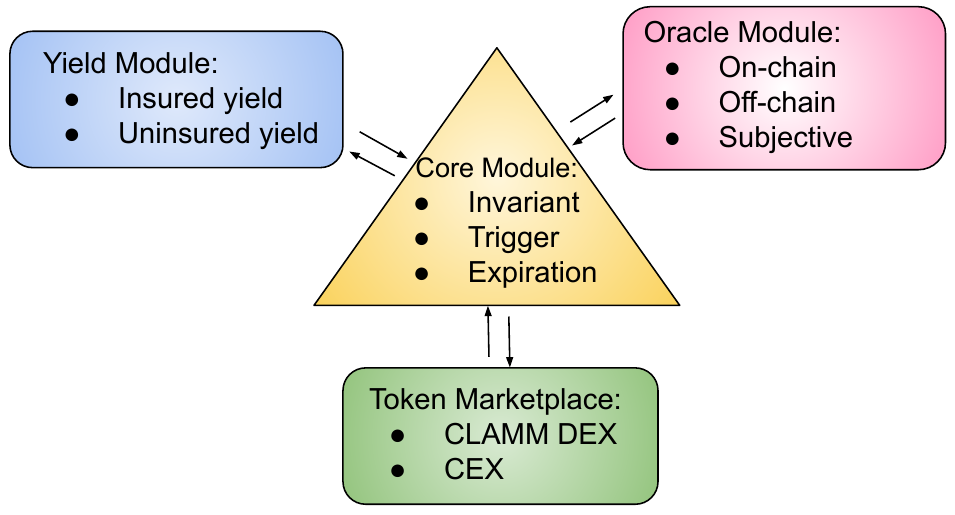}
\caption{An illustration of interactions within the {\name} protocol.}
\label{fig:backup-module}
\end{figure}
Consider an example where our protocol creates an insurance mechanism for the event of stETH de-pegging in which the price of stETH relative to ETH falls below $0.95$. 
In its simplest version, it creates two tokens: a policy token $\K$ and an underwriting token $\I$ for a predetermined period, \emph{e.g.,} $1$ year. The payoffs of these tokens depend on the occurrence of the underlying event: if the price falls below $0.95$ before the deadline, $\K$ pays $\$1$, while $\I$ pays $\$0$. Otherwise, after the deadline passes, $\K$ pays $\$0$, while $\I$ pays $\$1$. For $\K$-holders, it creates a hedging mechanism for the de-pegging event during the specified period, whereas $\I$-holders provide hedging in return for a premium.

The actual payoffs of $\K$ and $\I$ are slightly different from the above description, however, the underlying motivation remains the same. The protocol design is modular with key components divided into the core, oracle, and yield modules, as illustrated in Figure~\ref{fig:backup-module}. The subsequent subsections describe the modeling of financial adversities, followed by a detailed and formal description of each of these modules.

\subsection{Binary Event}
We define a \emph{binary event} being the tuple $(a,t,s)$ as follows: 
\begin{itemize}
\item Assertion $a$: A binary event is defined by an immutable assertion $a$. For the previous example, the assertion is defined as the price of stETH relative to ETH, $p_{\text{stETH/ETH}}$, being less than $0.95$ and represented as $p_{\text{stETH/ETH}}<0.95$.
    \item Expiration period $t$: This denotes the time duration since the inception of a binary event during which it remains valid. In the previous example, this was $1$ year.
    \item State $s$: A binary event can be in either of the two states: \emph{True} or \emph{False}. By default, an event is in the \emph{False} state. If the underlying assertion holds at any instance before the event's expiration, the state shifts to \emph{True} and becomes immutable. Moreover, the state does not change once the event expires.
\end{itemize}
Therefore, for an adverse effect modeled as a binary event, the occurrence of the $True$ state signifies an insurance claim.

\noindent \textbf{Modelling complex adversities:} A binary event only captures a discrete incident. However, in reality, adversities can have varying degrees of damage. To address this, our protocol requires an adversity to be modeled as a collection of 
$n\in \mathbb{N}$ binary events, where the $i^{th}$ event is represented as $(a_{i}, t, s_{i})$, with all of them having a common expiration.

For instance, a model of continuous incident of stETH de-pegging consists of the following three assertions:
\begin{enumerate}
    \item $a_{1}:$ $p_{\text{stETH/ETH}}<0.99$.
    \item $a_{2}:$ $p_{\text{stETH/ETH}}<0.98$.
    \item $a_{3}:$ $p_{\text{stETH/ETH}}<0.95$.
\end{enumerate}
In the above example, if only $a_1$ holds by the deadline, \emph{i.e.,} $(s_{1},s_{2},s_{3})=(True,False,False)$, it corresponds to a mild de-pegging incident; if both $a_1$ and $a_2$ hold, \emph{i.e.,} $(s_{1},s_{2},s_{3})=(True,True,False)$, it represents a moderate de-pegging incident; and lastly, if all the three assertions hold, \emph{i.e.,} $(s_{1},s_{2},s_{3})=(True,True,True)$, it represents an extreme de-pegging incident. Therefore, the final state values together signifies the degree of damage.
\subsection{Core Module}\label{sec:core-module}
\begin{figure}[t]
\includegraphics[width=9cm]{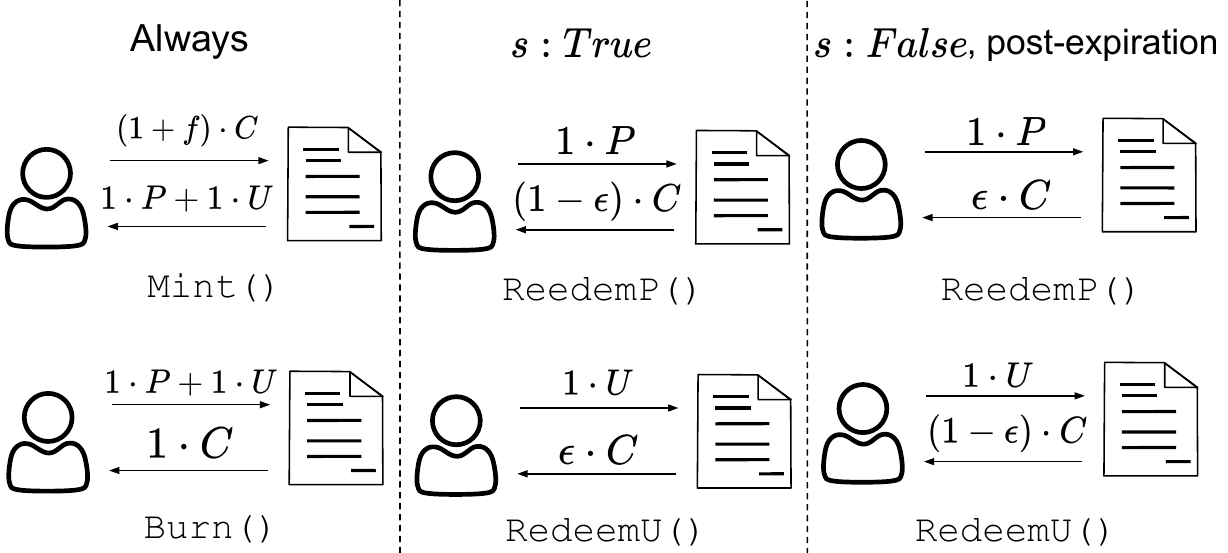}
\caption{Interaction between user and core module.}
\label{fig:core-module}
\end{figure}
For an adversity modeled as a collection of $n$ binary events, the core module contract creates $n$ pairs of ERC20 tokens $\K_{i},$ $\I_{i}$ with $i \in \{1\ldots n\}$. Each core contract is characterized by a base currency $\C$ (\emph{eg.,} ETH, USDC) in which claims are disbursed. The following module methods can be invoked for each binary event from the inception of the module to  perpetuity. Here $f$ represents a constant fee such that $f\in[0,1]$.
\begin{itemize}
    \item \texttt{Mint(i):} A user depositing $k(1+f)$ units of $\C$, where $k>0$, receives $k$ units each of $\K_{i}$ and $\I_{i}$. Here, $kf$ units of $\C$ get deposited as fees in the \texttt{feeTo} address specified by the core module's creator.
    \item \texttt{Burn(i):} A user depositing $k$ units of $\K_{i}$ and $\I_{i}$ receives $k$ units of $\C$.
\end{itemize}
These methods are depicted in the left part of Figure~\ref{fig:core-module} and they enforce the following two invariants:
\begin{invariant}\label{inv-1}
$(1+f)\cdot \C \Rightarrow \K_{i} + \I_{i}$
\end{invariant}
\begin{invariant}\label{inv-2}
$\K_{i} + \I_{i} \Rightarrow \C $  
\end{invariant}
Next, we describe methods to redeem the above tokens in exchange for $\C$ based on the state of a binary event. At any point in time, $\forall i$ s.t. $s_{i}=True$, $\K_{i}$ and $\I_{i}$ can be redeemed for the following:
\begin{equation}\label{eq:redeemTrue}
\K_{i}: (1-\epsilon)\cdot \C,\;\;\;\I_{i}: \epsilon\cdot \C
\end{equation}
Here $\epsilon$ is a constant whose value is set by the module creator and is generally close to zero, \emph{i.e.,} $0 < \epsilon \ll 1 $. The reason for such a choice of $\epsilon$ is justified in Section~\ref{sec:evaluation} as it relates to the risks of the LPs on AMM. After the events expire, the module tokens can be redeemed for binary events with $False$ state as well, \emph{i.e.,} $\forall i$ s.t. $s_{i}=False$, $\K_{i}$ and $\I_{i}$ can be redeemed for:
\begin{equation}\label{eq:redeemExpire}
\K_{i}: \epsilon\cdot \C,\;\;\;\I_{i}: (1-\epsilon)\cdot \C
\end{equation}
Note that the above payoff values satisfy \textbf{Invariant~\eqref{inv-2}} since the sum of payoffs of $1\cdot\K_{i}$ and $1\cdot \I_{i}$ is always $1\cdot\C$, regardless of the event's state. The redemptions are implemented using the \texttt{RedeemP(i)} and \texttt{RedeemU(i)} methods, respectively, which are depicted in the middle and right parts of Figure~\ref{fig:core-module}.

Let $p_{\K_i}, \: p_{\I_i}$ be the prices of $\K_{i}, \: \I_{i}$ in terms of $\C$. A policyholder purchasing $1 \cdot \K_{i}$ from the market pays a one-time premium of $p_{\K_{i}}$ to obtain an insurance coverage of $(1-\epsilon) \cdot \C$. This gives a coverage-to-premium ratio of the policy to be $(1-\epsilon)/p_{\K_{i}}$. 

On the other hand, a user underwrites a policy only if there is sufficient incentive for them. One such scenario is when the price of the policy token increases significantly. This is because \textbf{Invariant}~\eqref{inv-1},\eqref{inv-2} imply:
\begin{equation}\label{eq:p-u-price-rel}
    1\le p_{\K_{i}}+p_{\I_{i}}\le 1+f
\end{equation}
as one can always acquire $1 \cdot \K_{i}+1\cdot\I_{i}$ for $(1+f)\cdot C$ using \texttt{Mint()} and can always dispose the same for $1\cdot\C$ using \texttt{Burn()}. Therefore, if the price of the policy token increases by more than $f+\epsilon$, the following holds:
\begin{equation}
    f+\epsilon+p_{\I_{i}}<p_{\K_{i}}+p_{\I_{i}}\le 1+f
\end{equation}
Hence, the price of the underwriting token becomes less than $1-\epsilon$. This incentivizes underwriters to acquire $\I_{i}$ at a lower price and later redeem it for $(1-\epsilon)\cdot\C$ if the assertion does not hold, with the difference representing the earned premium.

In summary, the immutability of the core module and an invariant-based redemption guarantee zero risk of default, thereby ensuring the robustness of the {\name} protocol.

\subsection{Oracle Module}
\begin{figure}[t]
\includegraphics[width=9.5cm]{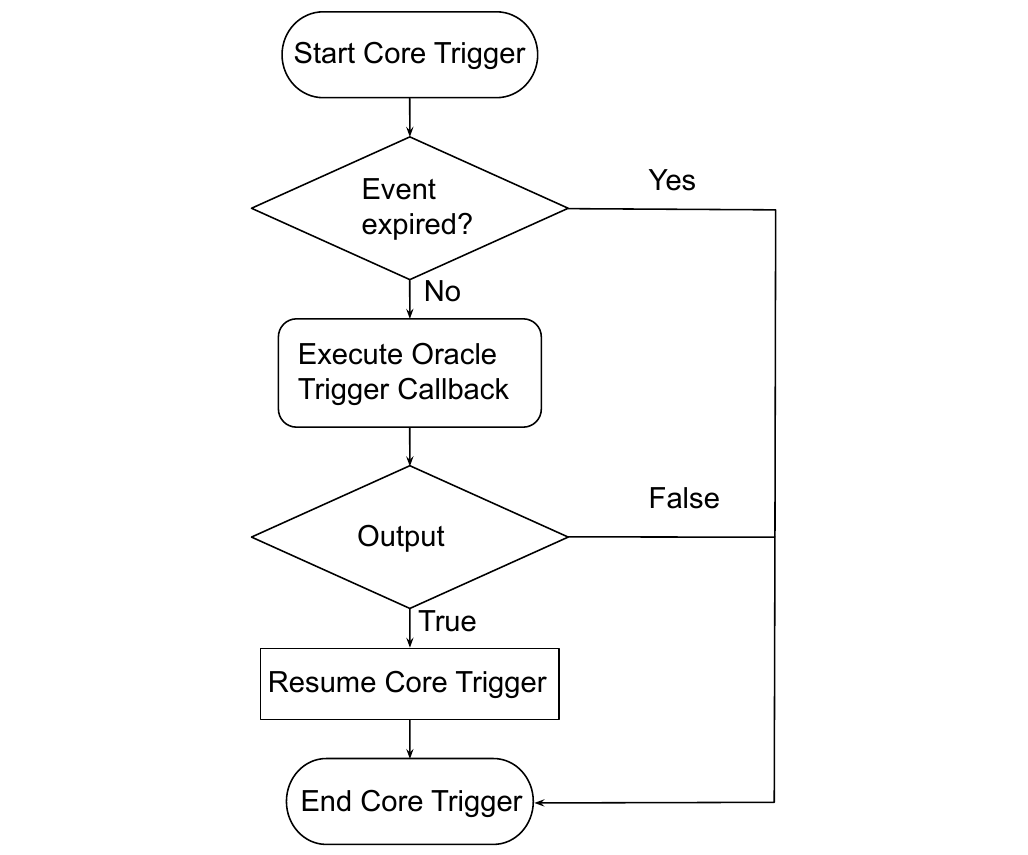}
\caption{Flow of Trigger execution using Oracle module.}
\label{fig:backup-oracle-flow}
\end{figure}

The oracle module is an externally deployed contract with a custom-defined logic. This logic takes input values from external data sources and executes the logic of assertion for each event. A user willing to create a core contract on {\name} specifies an oracle contract address to associate with it. This is because the core module consists of a \texttt{Trigger(i)} function for the $i^{th}$ event that executes a callback function \texttt{OTrigger(i)} in the oracle contract. \texttt{OTrigger(i)} executes the logic of the $i^{th}$ assertion and returns a boolean value back to \texttt{Trigger(i)} in the core contract. This is presented as a flowchart diagram in Figure~\ref{fig:backup-oracle-flow}.

The returned value $False$ is ignored; however, in the other case, the following occurs:
\begin{enumerate}
    \item $s_{i}$ transitions to \emph{True} and becomes immutable.
    \item The \texttt{RedeemP(i)}, and \texttt{RedeemU(i)} methods are set to exchange as per~\eqref{eq:redeemTrue}.
\end{enumerate}

When the contract deadline passes, the following occurs:
\begin{enumerate}
    \item The \texttt{Trigger()} method is rendered invalid.
    \item $s_{i}$ becomes immutable for all events.
    \item The \texttt{RedeemP(i)} and \texttt{RedeemU(i)} functions become valid for events with a $False$ state, redeeming tokens based on Relation~\eqref{eq:redeemExpire}.
\end{enumerate}

Custom developer-defined logic enables extensive functionalities for the oracle module. Simultaneously, a modular structure ensures that the accounting performed by the core module remains unaffected by the execution of the oracle module. 
This allows users to simultaneously reconcile platform resilience and flexibility of policies.

External information, such as prices and liquidity on AMM, can be sampled from on-chain or off-chain oracle sources. Given the potential vulnerability of oracle sources to manipulation~\cite{angeris2020improved,liu2021first}, a combination of sources with complementary strengths can be employed to mitigate such risks. We illustrate this with an example for the \texttt{OTrigger()} function below.\\

\noindent\textbf{Illustrative Example:} Listing~\ref{algo:oracle-code} presents  pseudocode for the \texttt{OTrigger()} function that detects a reduction in the price of stETH from ETH by more than $5\%$
. The algorithm achieves this by sampling data from two sources: (a) Uniswap V3 pool for stETH/ETH, an on-chain oracle source, and (b) Chainlink~\cite{breidenbach2021chainlink} price feed for stETH/ETH, an off-chain oracle source (without loss of generality for
the underlying price-feeding system). In the first, Uniswap provides the sum of the log of prices at each second since the inception of the pool until a given time. The algorithm samples these values at time $0$ and $3600$ seconds ago, and divides their difference by $3600$. This computation yields the log of time-weighted geometric mean of the price over the last $60$ minutes~\cite{adams2021uniswap}. If both this value and the price from Chainlink are less than the threshold\footnote{In Listing~\ref{algo:oracle-code}, the $log$ is computed with the base $1.0001$. Thus, \texttt{avgLogPrice} is compared with $log_{1.0001}(0.95)=-513$.}, \texttt{OTrigger()} returns $True$; otherwise, it returns $False$.

In the above example, it is in the interest of $\K$-holders to call \texttt{Trigger()} when de-pegging occurs. However, this incurs a gas fee to the caller, while the benefits of the call are shared among all $\K$-holders, creating a first-mover's disadvantage. To address this issue, the caller gets compensated with the fee collected by the \texttt{feeTo} address.
To detect uncommon events with insufficient or unreliable oracle sources, the oracle contract can delegate the assessment to decentralized dispute resolution platforms, such as Kleros~\cite{kleros}. This, coupled with the customizable nature of the oracle contract, broadens the range of events detectable by {\name}.

\begin{figure}[t]
\lstset{basicstyle=\ttfamily\footnotesize,
      stepnumber=1,numberblanklines=false,mathescape=true}
\begin{lstlisting}[numbersep=2pt]
fn OTrigger() returns (bool outcome) {
    address pool = 0xfff;
    seconds = [0,3600];
    logAgg = IUniswapV3Pool(pool).observe(seconds);
    logAggDelta = logAgg[1] -logAgg[0];
    avgLogPrice = logAggDelta / 3600;
    price2 = ChainlinkFeed.latestRoundData()
    if (avgLogPrice < -513 && price2 < 0.95){
        return True;
    } else {
        return False;
    }}
\end{lstlisting}
\vspace{-2.5mm}
\caption{A code snippet demonstrating oracle contract to detect de-pegging of stETH/ETH pair.}
\label{algo:oracle-code}
\vspace{-5mm}
\end{figure}

\subsection{Yield Module}\label{subsec:yield}
\begin{figure}[t]
\includegraphics[width=8.5cm]{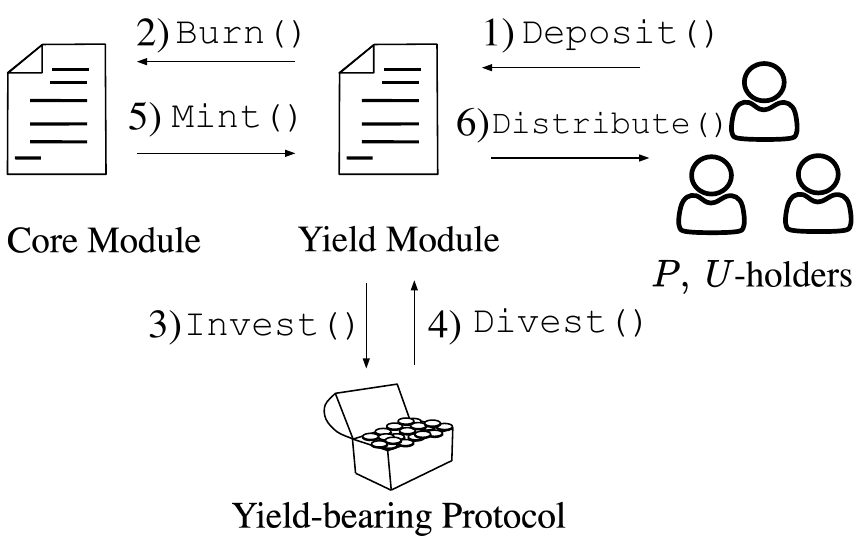}
\caption{Sequence of methods in the yield module with arrows indicating direction from caller to callee.}
\label{fig:yield-module}
\vspace{-4mm}
\end{figure}

The {\name} protocol enables custom peripheral yield modules where $\K$- and $\I$-holders have the \emph{option} to earn yield on their locked $\C$. 
The key idea is to utilize the perpetual nature of the \texttt{Mint()} and \texttt{Burn()} methods of the core module, along with the trustless execution of smart contracts. Below, we present methods for a module design that delegates yield generation to a DRT protocol comprising $A$- and $B$-tranches (low- and high-risk respectively), and earning yield on $\C$. Figure~\ref{fig:yield-module} depicts the sequence of the following methods.

\begin{enumerate}
    \item \texttt{Deposit():} $\K$- and $\I$-holders willing to earn yield within $A$-tranche of DRT deposit their tokens in the yield contract. Let there be $k_{\K}$ units of $\K$ and $k_{\I}$ units of $\I$ and let us define $\lambda$ as $min(k_{\K},k_{\I})$.
    \item \texttt{Invest():} Since one can only burn equal quantities of the two tokens, the yield module burns $\lambda$ units of $\K$ and $\I$ to obtain $\lambda$ units of $\C$ and invests this capital in DRT's $A$-tranche. Observe that for maximum utilization of the tokens, the quantities of the two tokens needs to be equal.
    \item \texttt{Divest():} After the expiration of DRT, the module divests its position from $A$-tranche to obtain $\lambda'$ units of $\C$. Based on the success of divesting, $\lambda'$ can be greater than, equal to, or less than $\lambda$ as described in Section~\ref{sec:DRT}.
    \item \texttt{Distribute():} The yield contract mints $\lambda'/(1+f)$ units of $\K$ and $\I$ and distributes  them, along with the unburnt tokens, uniformly among depositors.
\end{enumerate}
The above works analogously for the $B$-tranche of DRT.

Without loss of generality, let $k_{\K}\ge k_{\I}$ making $\lambda = k_{\I}$ and let $k'_{\I}=\lambda'/(1+f)$. Then, in the scenario of a successful divesting, $k'_{\I}-k_{\I}$ units of $\K$ and $\I$ are uniformly distributed as yield to their holders. Thus, the holder of $1\cdot \I$ receives a yield fraction of $(k'_{\I}-k_{\I})/k_{\I}$ while the holder of $1\cdot \K$ receives a yield fraction of $(k'_{\I}-k_{\I})/k_{\K}$, which is $k_{\I}/k_{\K}$ times the yield earned per unit of $\I$. Therefore, if there are more policyholders than underwriters, $k_{\I}/k_{\K}$ becomes smaller, attracting fewer additional policyholders than underwriters until $k_{\I}/k_{\K}$ returns to $1$. 

Optional yield earning and investing via DRT in the {\name} protocol naturally creates multiple categories of user
classification based on their risk appetite:
\begin{enumerate}
    \item \emph{High-risk / high-reward:} This encompasses underwriters investing in $B$-tranche. The risk for these users arises from the underlying insurance event, coupled with an increased divesting risk in the $B$-tranche of DRT. In return, they earn a premium for both underwriting insurance and investment in the $B$-tranche.   
    \item \emph{Medium-high risk / medium-high reward:} This includes:
\begin{itemize}
    \item Underwriters investing in $A$-tranche: They bear risk from underwriting, with a lower risk of divesting.
    \item Policyholders investing in $B$-tranche: Their risk related to adversity is hedged, but they face a higher risk of divesting.
\end{itemize}

    \item \emph{Medium-low risk / medium-low reward:} This includes:
\begin{itemize}
    \item Underwriters not investing at all: They bear risk solely from underwriting.
    \item Policyholders investing in $A$-tranche: Their risk related to adversity is hedged, and they face a lower risk of divesting.
\end{itemize}
    \item \emph{Low-risk / Low-reward:} This includes policyholders not investing at all. While they are hedged against adversity with no additional risk, they miss out on yield returns.
\end{enumerate}

Since yield-bearing smart contract protocols are susceptible to vulnerabilities, the peripheral design of the yield module ensures that it doesn't disrupt the functionality of the other two modules. Moreover, the above methods can be adapted, allowing room for custom designs and enabling users to earn yield on their preferred platform.

In summary, instead of making decisions on the user's behalf, the above design grants them the flexibility to make individualized risk management decisions, {\em i.e.,} divesting early or substituting the underlying yield-bearing platform if possible, without impacting others. 
While having flexible policies, users therefore trade-off between platform risk exposure and capital efficiency.

\section{Token Marketplace}\label{sec:marketplace}

For the proper functioning of {\name}, it is essential to have an efficient marketplace where users can buy or sell policy and underwriting tokens in exchange for a common currency. Therefore, studying the incentive mechanism for various users to participate in the marketplace is crucial. This includes examining the risk-adjusted returns for the liquidity providers. In this paper, we focus on studying LP strategies for the trading pair $\K_{i}/\I_{i}$ on AMMs, although other pairs such as $\K_{i}/\C$ or $\I_{i}/\C$ can also exist. This is because the former pair is sufficient for buying or selling $\K_{i}$ and $\I_{i}$ in exchange for $\C$ (as explained next). Another advantage is that the liquidity provision for this trading pair does not require currency tokens.

Let $p_{i}$ denote the spot price of $\K_{i}$ relative to $\I_{i}$. A user purchasing a small quantity $\delta$ of $\K_{i}$ (such that slippage on AMM is negligible) can perform the following ordered actions in a single transaction:
\begin{itemize}
    \item Mint $\frac{\delta p_{i}}{1+p_{i}}$ units of $\K_{i}$ and $\I_{i}$ by depositing $\frac{\delta p_{i}}{1+p_{i}}(1+f)\cdot\C$.
    \item Swap $\frac{\delta p_{i}}{1+ p_{i}}\cdot\I_{i}$ for $\frac{\delta}{1+p_{i}}\cdot\K_{i}$ on AMM.
\end{itemize}
The user now holds $\delta\cdot\K_{i}$. Similar can be done when the user wants to sell $\delta\cdot\K_{i}$:
\begin{itemize}
    \item Swap $\frac{\delta}{p_{i}+1}\cdot\K_{i}$ for $\frac{\delta p_{i}}{1+p_{i}}\cdot\I_{i}$ on AMM.
    \item Burn equal $\K_{i}$ and $\I_{i}$ to receive $\frac{\delta p_{i}}{1+p_{i}}\cdot\C$.
\end{itemize}
The procedure for buying and selling underwriting tokens is analogous to the above. As a result, the effective buying and selling prices of $\K_{i}$ in terms of $\C$ are $\frac{p_{i}}{1+p_{i}}(1+f)$ and $\frac{p_{i}}{1+p_{i}}$, respectively and for $\I_{i}$, they are $\frac{1}{1+p_{i}}(1+f)$ and $\frac{1}{1+p_{i}}$, respectively. Therefore, the market's risk assessment results in different values of policy and underwriting tokens, and it gets expressed as a single price $p_{i}$. The following describes various strategies for liquidity providers on concentrated liquidity AMMs, which we then evaluate in the subsequent section.\\

\noindent\textbf{LP strategy}: Assuming $f\approx0$ and negligible risk-free interest rate, the value of $p_{i}$ satisfies:
\begin{equation}\label{eq:pi-price-ineq}
\frac{\epsilon}{1-\epsilon}\le p_{i} \le \frac{1-\epsilon}{\epsilon}
\end{equation}
This is because if $p_{i}<\frac{\epsilon}{1-\epsilon}$, then the price of $\K_{i}$ relative to $\C$ satisfies:
\begin{equation}
    p_{\K_{i}}=\frac{p_{i}}{1+p_{i}} = (1- \frac{1}{1+p_{i}}) < \epsilon
\end{equation}
Given that $\K_{i}$ can be redeemed for at least $\epsilon\cdot\C$ at the contract expiration, irrespective of the final state, a trader can purchase them at a price $p_{\K_{i}}$ and subsequently redeem them at a higher price of $\epsilon$ at contract expiration, thus creating an arbitrage opportunity.\\


Similarly, if $p_{i}>\frac{1-\epsilon}{\epsilon}$, then the price of $\I_{i}$ relative to $\C$ satisfies:
\begin{equation}
    p_{\I_{i}} = (\frac{1}{1+p_{i}}) < \epsilon
\end{equation}
As $\I_{i}$ can also be redeemed for at least $\epsilon\cdot\C$ upon contract expiration, another arbitrage opportunity arises. A trader can purchase them at a lower price of $p_{\I_{i}}$ and subsequently redeem them for at least $\epsilon\cdot\C$ at the contract expiration. Consequently, Inequality~\eqref{eq:pi-price-ineq} holds.

\begin{figure}[!t]
\centering
\subfloat[Uniform LP strategy]{\includegraphics[width=3in]{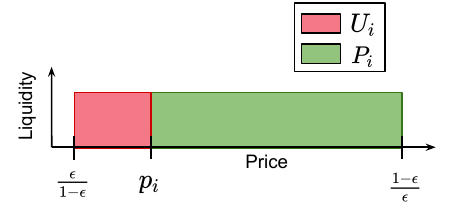}
\label{subfig:lp-a}}

\subfloat[$p_{i}$ concentrated LP strategy]{\includegraphics[width=3in]{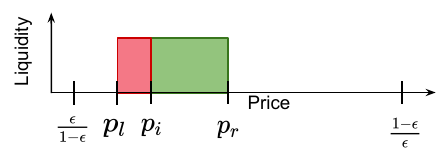}
\label{subfig:lp-b}}

\subfloat[Edge concentrated LP strategy]{\includegraphics[width=3in]{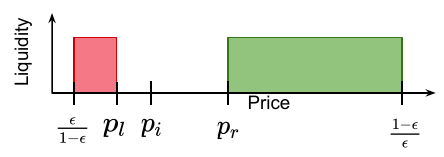}
\label{subfig:lp-c}}
\caption{Liquidity provision under the three categories.}
\label{fig_sim}
\vspace{-4mm}
\end{figure}

Considering a concentrated liquidity pool for the pair $\K_{i}/\I_{i}$, the LPs thus need to deposit policy tokens in the price interval $[p_{i}, \frac{1-\epsilon}{\epsilon}]$ (prices above the spot price) 
and underwriting tokens in the price interval $[\frac{\epsilon}{1-\epsilon}, p_{i}]$ (prices below the spot price) with some distribution. 
Suppose an LP mints $1$ unit each of two tokens, and provisions them as liquidity with $p_{i}$ being the entry spot price. We consider three limiting passive strategies for liquidity provision:
\begin{enumerate}
    \item \textbf{Uniform:} This strategy consists of depositing $1\cdot \K_{i}$ in the price interval $[p_{i},\frac{1-\epsilon}{\epsilon}]$, and $1\cdot \I_{i}$ in the price interval $[\frac{\epsilon}{1-\epsilon},p_{i}]$ with the two intervals having a constant liquidity. This is depicted in Figure~\ref{subfig:lp-a}.
    \item \textbf{Concentrated around $p_{i}$:} This strategy involves depositing, with a constant liquidity, $1\cdot \K_{i}$ in $[p_{i},p_{r}]$, and $1\cdot \I_{i}$ in $[p_{l},p_{i}]$ for some $p_{l},p_{r}$. Such a strategy is expected to earn a higher trading fee than the uniform strategy since liquidity is concentrated around the spot price. This is depicted in Figure~\ref{subfig:lp-b}.
    
    \item \textbf{Concentrated at the edges:} This strategy consists of depositing, with a constant liquidity, $1\cdot \K_{i}$ in $[p_{r}, \frac{1-\epsilon}{\epsilon}]$, and $1\cdot \I_{i}$ in $[\frac{\epsilon}{1-\epsilon},p_{l}]$ for some $p_{l},p_{r}$. Such a strategy is expected to earn lower trading fees than the uniform strategy since liquidity is concentrated away from the spot price. This is depicted in Figure~\ref{subfig:lp-c}.
\end{enumerate}
The next Section evaluates the divergence loss of the LPs under the above strategies and examines ways to mitigate risk.

\section{Divergence Loss Evaluation}\label{sec:evaluation}
\begin{figure*}[t]
     \centering
     \begin{subfigure}[b]{0.32\textwidth}
         \centering
         \includegraphics[width=\textwidth]{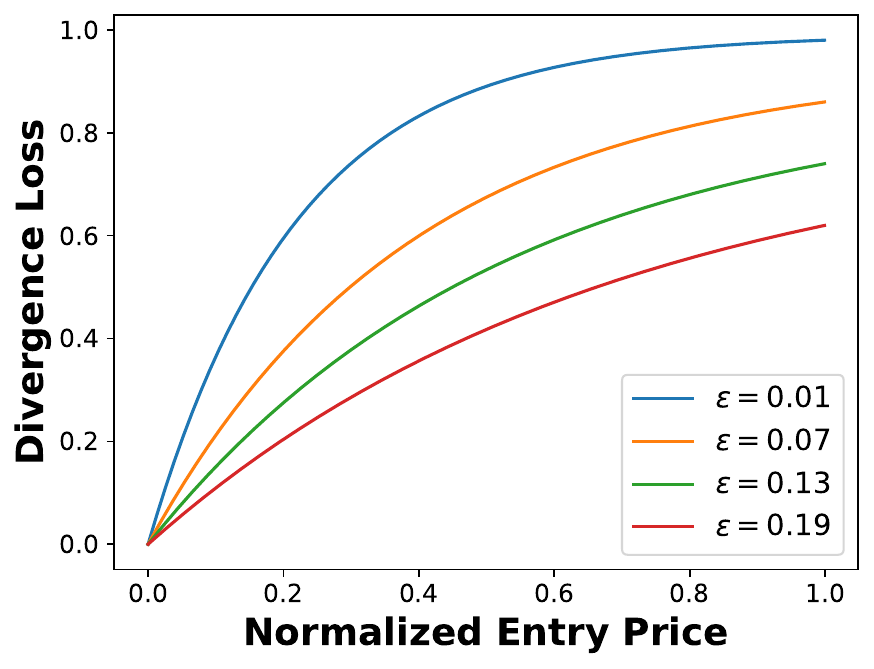}
         \caption{No catastrophe, uniform strategy}
         \label{fig:uni-opt}
     \end{subfigure}
     \hfill
     \begin{subfigure}[b]{0.32\textwidth}
         \centering
         \includegraphics[width=\textwidth]{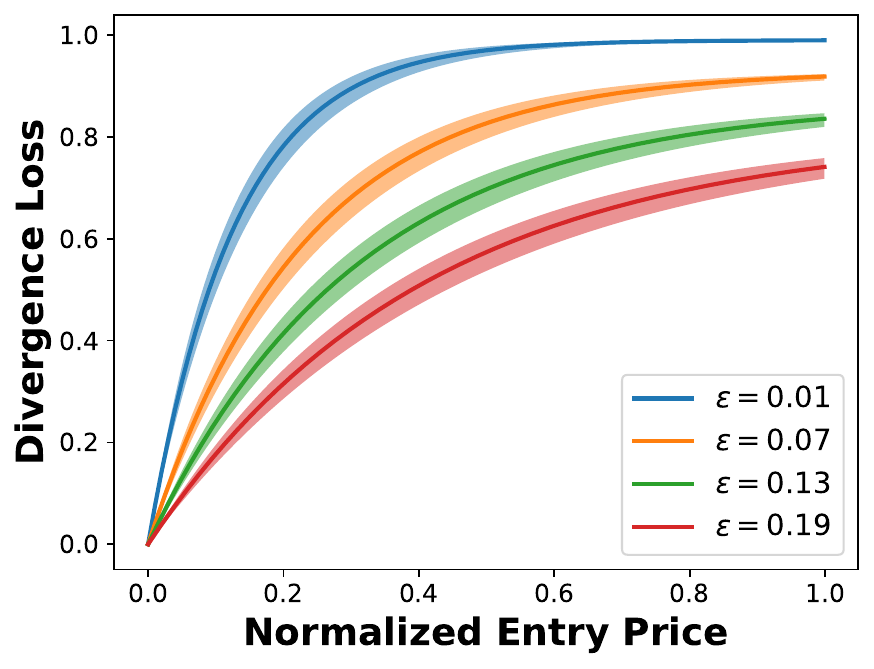}
         \caption{No catastrophe, $\textbf{p}$ concentrated strategy}
         \label{fig:conc-opt}
     \end{subfigure}
     \hfill
     \begin{subfigure}[b]{0.32\textwidth}
         \centering
         \includegraphics[width=\textwidth]{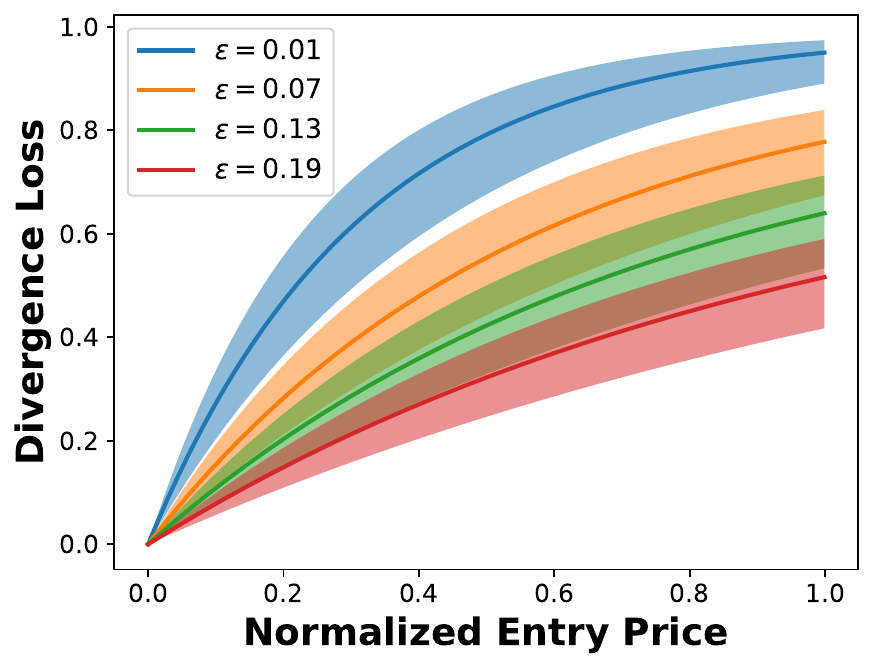}
         \caption{No catastrophe, edge strategy}
         \label{fig:edge-opt}
     \end{subfigure}
     \hfill
     \begin{subfigure}[b]{0.32\textwidth}
         \centering
         \includegraphics[width=\textwidth]{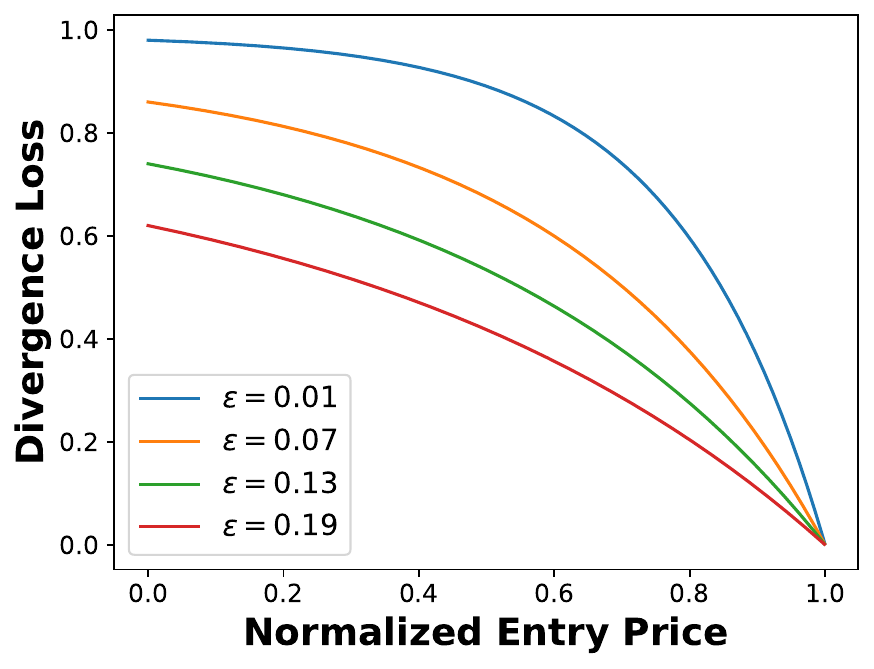}
         \caption{Catastrophe, uniform strategy}
         \label{fig:uni-pess}
     \end{subfigure}
     \hfill
     \begin{subfigure}[b]{0.32\textwidth}
         \centering
         \includegraphics[width=\textwidth]{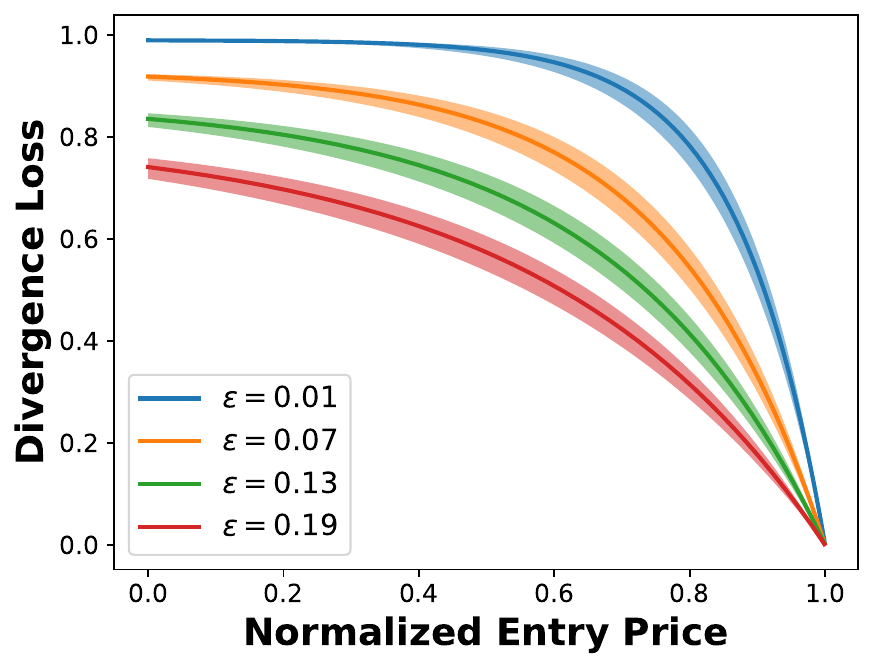}
         \caption{Catastrophe, $\textbf{p}$ concentrated strategy}
         \label{fig:conc-pess}
     \end{subfigure}
     \hfill
     \begin{subfigure}[b]{0.32\textwidth}
         \centering
         \includegraphics[width=\textwidth]{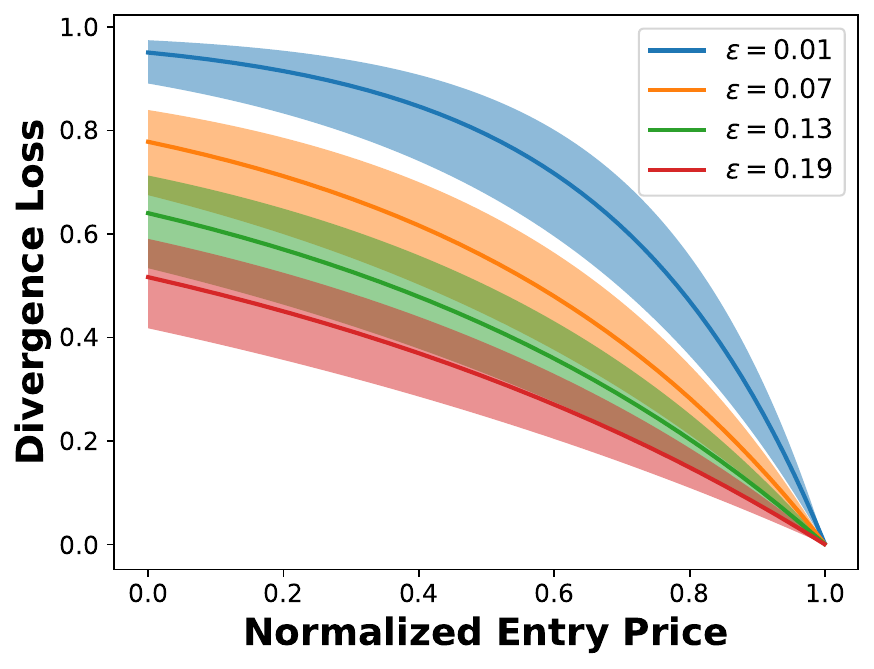}
         \caption{Catastrophe, edge strategy}
         \label{fig:edge-pess}
     \end{subfigure}
        \caption{Comparison of divergence loss among three liquidity provision strategies (uniform, concentrated around the entry price, concentrated at the interval edges) in scenarios without and with catastrophes.}
        \label{lp-loss}
\end{figure*}
\emph{Divergence Loss} is defined as the opportunity cost for
an LP to provide token reserves as liquidity compared to just holding them. In this section, we assess the divergence loss experienced by liquidity providers across the three strategies: uniform, concentrated around the entry price, and concentrated at the price interval edges (Figure~\ref{fig_sim}). Our evaluation addresses the following two questions: $(1)$ Which parameters significantly impact the divergence loss of liquidity providers and to what extent? $(2)$ What are some optimal techniques to mitigate the risk of divergence loss for liquidity providers?

We adhere to an evaluation methodology where we assume that an LP mints $1$ unit of $\K_{i}$ and $\I_{i}$ and creates a liquidity position with an entry price $p_{i}=\textbf{p}$. If there is no liquidity provision, the value of these tokens at the conclusion of the contract is always $1$ unit of $\C$ (enforced by Invariant~\eqref{inv-2}). To evaluate the divergence loss, we calculate the total value of the tokens held by the LP at the conclusion of the contract and represent it as a fraction of $1\cdot\C$. We use four equally-spaced values for $\epsilon:$ $\{0.01, 0.07, 0.13, 0.19\}$ and do not consider any further higher values since $\epsilon\ll1$. For plotting, we take the natural logarithm of the x-axis, for the price interval $[\frac{\epsilon}{1-\epsilon}, \frac{1-\epsilon}{\epsilon}]$, and then min-max normalize it to the range $[0,1]$. Here is a summary of divergence loss in the three strategies:
\begin{enumerate}
    \item Uniform: If there is no catastrophe, \emph{i.e.,} the underlying assertion does not hold, $p_{i}=\frac{\epsilon}{1-\epsilon}$ at event expiration and the LP ends up with only $\K_{i}$. Their $1\cdot\I_{i}$ is exchanged for an average price of $\sqrt{\frac{\epsilon \textbf{p}}{1-\epsilon}}$, which is the geometric mean of the extreme prices $\{\frac{\epsilon}{1-\epsilon},\textbf{p}\}$ as derived in Section~\ref{sec:conc-liq}. Therefore, the LP ends up with $1+\sqrt{\frac{1-\epsilon}{\epsilon \textbf{p}}}$ units of $\K_{i}$ each worth $\epsilon\cdot\C$, thus a total value of $\epsilon+\sqrt{\frac{\epsilon(1-\epsilon)}{\textbf{p}}}$ units of $\C$.

    When there is a catastrophe, the LP ends up with only $\I_{i}$. As above, their $1\cdot\K_{i}$ is exchanged for an average price of $\sqrt{\frac{(1-\epsilon) \textbf{p}}{\epsilon}}$, giving them a total of $1+\sqrt{\frac{(1-\epsilon) \textbf{p}}{\epsilon}}$ units of $\I_{i}$. Since $\I_{i}$ is worth $\epsilon\cdot\C$, they receive a value of $\epsilon+\sqrt{\textbf{p}\epsilon(1-\epsilon)}$ units of $\C$.

    The above two cases are shown in Figure~\ref{fig:uni-opt} \&~\ref{fig:uni-pess}, respectively. We can observe that when the normalized entry price is closer to $0$$(1)$, the divergence loss is higher for the catastrophic(non-catastrophic) scenario. 
    Also, for a given $\textbf{p}$, the divergence loss is lower at higher values of $\epsilon$. The worst loss (at \textbf{p}=$\frac{\epsilon}{1-\epsilon}$ or $\frac{1-\epsilon}{\epsilon}$) can be reduced from $0.98$ to $0.62$ by varying $\epsilon$ from $0.01$ to $0.19$.
    
    \item Concentrated around $\textbf{p}$: In the case with no catastrophe, the final price $\frac{\epsilon}{1-\epsilon}$ is less than $p_{l}$. Therefore, the LP's $1\cdot\I_{i}$ is exchanged for an average price of $\sqrt{p_{l}\textbf{p}}$, which is the geometric mean of the extreme prices $\{p_{l},\textbf{p}\}$. As before, this gives a total value of $\epsilon+\frac{\epsilon}{\sqrt{p_{l}\textbf{p}}}$ units of $\C$. If there is a catastrophe, 
    the LP's $1\cdot\K_{i}$ is exchanged for an average price of $\sqrt{\textbf{p}p_{r}}$. This gives a total value of $\epsilon+\epsilon\sqrt{\textbf{p}p_{r}}$ units of $\C$.

    These two cases are shown in Figure~\ref{fig:conc-opt} \&~\ref{fig:conc-pess}. For a given $\textbf{p}$ and $\epsilon$, we choose a  range of values for the widths $\textbf{p}-p_{l},$ and $p_{r}-\textbf{p}$. This range represents $10-50\%$ of $\textbf{p}$ for $\textbf{p}-p_{l}$, and $10-50\%$ of $1-\textbf{p}$ for $p_{r}-\textbf{p}$ and is represented by the shaded region in the figure. The median value of the widths, given by $\textbf{p}-p_{l}=0.3\textbf{p}$, and $p_{r}-\textbf{p}=0.3(1-\textbf{p})$, is represented by the solid line. For a given $\textbf{p}$ and $\epsilon$, the divergence loss in this strategy is higher compared to the uniform strategy, particularly between $1\text{-}19.4\%$ at the interval edges (\textbf{p}=$\frac{\epsilon}{1-\epsilon}$ or $\frac{1-\epsilon}{\epsilon}$) for different values of $\epsilon$, and median values for $p_{l}$ and $p_{r}$. This is the cost LPs need to pay to achieve capital concentration (hence a higher fee revenue) in this strategy. However, the LPs can reduce their loss by choosing $p_{l},$ $p_{r}$ farther from $\textbf{p}$. The divergence loss, as in the previous case, reduces for higher values of $\epsilon$ with the worst loss reducing from $0.99$ to $0.74$ by choosing $\epsilon$ from $0.01$ to $0.19$.
    
    \item Concentrated at the edges: As before, in the case with no catastrophe, the LP's $1\cdot\I_{i}$ is exchanged for an average price of $\sqrt{\frac{\epsilon p_{l}}{1-\epsilon}}$. This gives a total value of $\epsilon+\sqrt{\frac{\epsilon(1-\epsilon)}{ p_{l}}}$ units of $\C$. If there is a catastrophe, the LP's $1\cdot\K_{i}$ is exchanged for an average price of $\sqrt{\frac{p_{r}(1-\epsilon)}{\epsilon}}$. This gives a total value of $\epsilon+\sqrt{p_{r}\epsilon(1-\epsilon)}$ units of $\C$.
    These cases are shown in Figure~\ref{fig:edge-opt} \&~\ref{fig:edge-pess} respectively. For a given $\textbf{p}$ and $\epsilon$, we chose a  range of values for the widths $\textbf{p}-p_{l},$ and $p_{r}-\textbf{p}$ as before. One can observe that the divergence loss in this strategy is the least of the three strategies. Compared to the uniform strategy, it is lower by $3.1\text{-}16.1\%$ at the interval edges (\textbf{p}=$\frac{\epsilon}{1-\epsilon}$ or $\frac{1-\epsilon}{\epsilon}$), and median values for $p_{l}$ and $p_{r}$. The loss increases when $p_{l},$ $p_{r}$ are farther from the edges. The lower divergence loss comes from a lower expected fee revenue since the position remains inactive when the price is outside its range. Lastly, the divergence loss reduces further with higher values of $\epsilon$, with the worst loss reducing from $0.95$ to $0.51$ by choosing $\epsilon$ from $0.01$ to $0.19$. This reduction is more than $47\%$ when compared to uniform LP strategy with $\epsilon=0.01$.
\end{enumerate}
Of the two tokens (policy and underwriting) that an LP initially holds, they always end up with the token of lesser value. If $\epsilon$ is set to $0$, the LP ends with tokens worth zero and incurs a divergence loss of $1$. As this is unfavourable to LPs, it underscores the necessity of enforcing $\epsilon>0$.

The three LP strategies present a {\em  trade-off} between divergence loss and increased fee revenue, resulting from enhanced capital concentration near the entry price. Additionally, an entry price closer to $0$ entails lower risk in a non-catastrophic outcome but higher risk in a catastrophic outcome, while the opposite is true for an entry price closer to $1$. Opting for a larger value of $\epsilon$ is another way to mitigate divergence loss risk, bringing it to as low as $0.51$ in the worst-case. However, since the policy tokens pays at least $\epsilon\cdot\C$ regardless of the final state, one consequence of larger $\epsilon$ is a higher token price relative to $\C$. This results in a lower coverage-to-premium ratio ($(1-\epsilon)/p_{\K_{i}}$), even for low-probability adversities, and thus, one must consider this trade-off.


To summarize, {\name} gives its users--policyholders, underwriters, and liquidity providers--the tools to individually manage their risks and rewards.  This is achieved through a modular approach. The invariant-based and immutable design ensures the robustness of the core module, the customizability of the oracle module enforces the flexibility of the insured policy, and most significantly, the optional and peripheral yield module empowers users to trade-off capital efficiency with external risk without compromising the security of the other users. 
\section{Related work}\label{sec:related}
\noindent\textbf{Prediction Markets:} These markets involve payoff contracts contingent on future events. Specifically, a contract is termed \emph{winner-take-all} if priced at $p$ units of $\C$, 
its payoff is $1\cdot\C$ only if a specific event occurs. The price of such contracts reflect the market's expectation of the probability of the contingent event occurring, assuming risk neutrality~\cite{wolfers2004prediction}. For the case where $\epsilon=0$, the payoff of the {\name}'s policy token aligns with that of a winner-take-all contract contingent on the underlying assertion of the binary event.


\noindent\textbf{Decentralized Insurance:} There have been several designs proposed for smart contract-based decentralized insurance for DeFi. Nexus Mutual~\cite{nexusMutual} and inSure~\cite{insure} are the largest of them and are inspired by the design of mutual insurance where platform users have a stake in the commercial success of the protocol~\cite{albrecht2017fundamental}. Unlike the work presented here, they do not support permissionless policy listing or optional yield generation. Rather, a fraction of the pooled capital is invested in yield-generation protocols, transferring risk to all members. These decisions and others, like claim assessment, are done via voting using governance tokens. However, in the case of Nexus Mutual, the majority of NXM tokens are held by a handful of addresses.\footnote{https://dune.com/queries/1445980/2529493} 

Other protocols, including ours, are based on Peer-to-Peer decentralized insurance~\cite{feng2022unified} where individuals pool their insurance premiums and use these funds to mitigate individual damages. Some of them including Risk Harbour~\cite{riskHarbor}, Opium Protocol~\cite{opium}, cozy.finance~\cite{cozy}, and Etherisc~\cite{etherisc} facilitate automated underwriting done via smart contract while others like Unslashed Finance~\cite{unslash}, Nsure~\cite{nsure}, and Cover~\cite{cover} have a voting-based claim assessment. Moreover, some of them including~\cite{opium,cozy} allow permissionless listing. In contrast to {\name}, none of the above protocols give their users the optional ability to manage yield on their capital. Instead, either the protocol offers no yield, or the yield management is performed by governance or delegated to a third party~\cite{enzyme}. Some designs, including Risk Harbour V1~\cite{riskHarborv1}, propose using receipt tokens from yield platforms (\emph{e.g.,} cTokens on Compound) as disbursement currency. Such an approach permanently links the risk of the yield platform with the protocol's underwriting module. On the other hand, {\name} disjoints the core, oracle, and yield modules and allows users to manage yield on their staked capital while having a permissionless policy creation.

\noindent\textbf{}
\section{Conclusion}\label{sec:conclusion}
A robust risk management system for the emerging area of DeFi requires a primitive platform with minimal external dependencies. This is successfully achieved in the {\name} protocol presented in this paper through a modular approach. At the same time, simulations suggest various mitigation strategies to minimize the risk for liquidity providers, contributing to the establishment of an efficient AMM marketplace. Possible future work in this direction
includes \emph{(a)} extending the conditional tokens from binary to $n$-ary with $n$ states to reduce the number of ERC20 tokens per adversity; \emph{(b)} enabling the creation of a single underwriting and multiple corresponding policy tokens to model a collection of adversities as a single event. Such an approach is expected to reduce the total capital requirement from underwriters, thus improving capital efficiency. 

\bibliographystyle{IEEEtran}
\bibliography{main}

\begin{thebibliography}{10}
\providecommand{\url}[1]{#1}
\csname url@samestyle\endcsname
\providecommand{\newblock}{\relax}
\providecommand{\bibinfo}[2]{#2}
\providecommand{\BIBentrySTDinterwordspacing}{\spaceskip=0pt\relax}
\providecommand{\BIBentryALTinterwordstretchfactor}{4}
\providecommand{\BIBentryALTinterwordspacing}{\spaceskip=\fontdimen2\font plus
\BIBentryALTinterwordstretchfactor\fontdimen3\font minus \fontdimen4\font\relax}
\providecommand{\BIBforeignlanguage}[2]{{%
\expandafter\ifx\csname l@#1\endcsname\relax
\typeout{** WARNING: IEEEtran.bst: No hyphenation pattern has been}%
\typeout{** loaded for the language `#1'. Using the pattern for}%
\typeout{** the default language instead.}%
\else
\language=\csname l@#1\endcsname
\fi
#2}}
\providecommand{\BIBdecl}{\relax}
\BIBdecl

\bibitem{nakamoto2008peer}
S.~Nakamoto, ``Bitcoin: A peer-to-peer electronic cash system,'' 2008, \url{https://bitcoin. org/bitcoin. pdf}.

\bibitem{buterin2014ethereum}
V.~Buterin, ``Ethereum: A next-generation smart contract and decentralized application platform,'' 2014, \url{https://ethereum.org/en/whitepaper/}.

\bibitem{avalanche}
T.~Rocket, ``Snowflake to avalanche: A novel metastable consensus protocol family for cryptocurrencies,'' \emph{Available [online].[Accessed: 4-12-2018]}, 2018.

\bibitem{yakovenko2018solana}
A.~Yakovenko, ``Solana: A new architecture for a high performance blockchain v0. 8.13,'' \emph{Whitepaper}, 2018.

\bibitem{nightshade}
A.~Skidanov and I.~Polosukhin, ``{Nightshade: Near Protocol Sharding Design},'' 2019, \url{https://near.org/downloads/Nightshade.pdf}.

\bibitem{defiVolume}
H.~Arslanian, ``Decentralised finance (defi),'' in \emph{The Book of Crypto: The Complete Guide to Understanding Bitcoin, Cryptocurrencies and Digital Assets}.\hskip 1em plus 0.5em minus 0.4em\relax Springer, 2022, pp. 291--313.

\bibitem{atzei2017survey}
N.~Atzei, M.~Bartoletti, and T.~Cimoli, ``A survey of attacks on ethereum smart contracts (sok),'' in \emph{Principles of Security and Trust: 6th International Conference, POST 2017, Held as Part of the European Joint Conferences on Theory and Practice of Software, ETAPS 2017, Uppsala, Sweden, April 22-29, 2017, Proceedings 6}.\hskip 1em plus 0.5em minus 0.4em\relax Springer, 2017, pp. 164--186.

\bibitem{gudgeon2020decentralized}
L.~Gudgeon, D.~Perez, D.~Harz, B.~Livshits, and A.~Gervais, ``The decentralized financial crisis,'' in \emph{2020 crypto valley conference on blockchain technology (CVCBT)}.\hskip 1em plus 0.5em minus 0.4em\relax IEEE, 2020, pp. 1--15.

\bibitem{zhou2023sok}
L.~Zhou, X.~Xiong, J.~Ernstberger, S.~Chaliasos, Z.~Wang, Y.~Wang, K.~Qin, R.~Wattenhofer, D.~Song, and A.~Gervais, ``Sok: Decentralized finance (defi) attacks,'' in \emph{2023 IEEE Symposium on Security and Privacy (SP)}.\hskip 1em plus 0.5em minus 0.4em\relax IEEE, 2023, pp. 2444--2461.

\bibitem{werner2022sok}
S.~Werner, D.~Perez, L.~Gudgeon, A.~Klages-Mundt, D.~Harz, and W.~Knottenbelt, ``Sok: Decentralized finance (defi),'' in \emph{Proceedings of the 4th ACM Conference on Advances in Financial Technologies}, 2022, pp. 30--46.

\bibitem{bekemeier2021deceptive}
F.~Bekemeier, ``Deceptive assurance? a conceptual view on systemic risk in decentralized finance (defi),'' in \emph{Proceedings of the 2021 4th International Conference on Blockchain Technology and Applications}, 2021, pp. 76--87.

\bibitem{lido2020}
``Lido: Ethereum liquid staking,'' 2020, \url{https://lido.fi/static/Lido:Ethereum-Liquid-Staking.pdf}.

\bibitem{nadler2023defi}
M.~Nadler, F.~Bekemeier, and F.~Sch{\"a}r, ``Defi risk transfer: Towards a fully decentralized insurance protocol,'' in \emph{2023 IEEE international conference on blockchain and cryptocurrency (ICBC)}.\hskip 1em plus 0.5em minus 0.4em\relax IEEE, 2023, pp. 1--9.

\bibitem{aave}
``Aave v3 technical paper,'' 2022, \url{https://github.com/aave/aave-v3-core/blob/master/techpaper/}.

\bibitem{leshner2019compound}
R.~Leshner and G.~Hayes, ``Compound: The money market protocol,'' \emph{White Paper}, 2019.

\bibitem{duffie2001risk}
D.~Duffie and N.~Garleanu, ``Risk and valuation of collateralized debt obligations,'' \emph{Financial analysts journal}, vol.~57, no.~1, pp. 41--59, 2001.

\bibitem{adams2021uniswap}
H.~Adams, N.~Zinsmeister, M.~Salem, R.~Keefer, and D.~Robinson, ``Uniswap v3 core,'' \emph{Tech. rep., Uniswap, Tech. Rep.}, 2021.

\bibitem{loesch2021impermanent}
S.~Loesch, N.~Hindman, M.~B. Richardson, and N.~Welch, ``Impermanent loss in uniswap v3,'' \emph{arXiv preprint arXiv:2111.09192}, 2021.

\bibitem{angeris2020improved}
G.~Angeris and T.~Chitra, ``Improved price oracles: Constant function market makers,'' in \emph{Proceedings of the 2nd ACM Conference on Advances in Financial Technologies}, 2020, pp. 80--91.

\bibitem{liu2021first}
B.~Liu, P.~Szalachowski, and J.~Zhou, ``A first look into defi oracles,'' in \emph{2021 IEEE International Conference on Decentralized Applications and Infrastructures (DAPPS)}.\hskip 1em plus 0.5em minus 0.4em\relax IEEE, 2021, pp. 39--48.

\bibitem{breidenbach2021chainlink}
L.~Breidenbach, C.~Cachin, B.~Chan, A.~Coventry, S.~Ellis, A.~Juels, F.~Koushanfar, A.~Miller, B.~Magauran, D.~Moroz \emph{et~al.}, ``Chainlink 2.0: Next steps in the evolution of decentralized oracle networks,'' \emph{Chainlink Labs}, vol.~1, pp. 1--136, 2021.

\bibitem{kleros}
W.~G. Clement~Lesaege and F.~Ast, ``Kleros:a decentralized decision protocol,'' 2021, \url{https://kleros.io/yellowpaper.pdf}.

\bibitem{wolfers2004prediction}
J.~Wolfers and E.~Zitzewitz, ``Prediction markets,'' \emph{Journal of economic perspectives}, vol.~18, no.~2, pp. 107--126, 2004.

\bibitem{nexusMutual}
H.~Karp and R.~Melbardis, ``Nexus mutual whitepaper: A peer-to-peer discretionary mutual on the ethereum blockchain,'' 2017, \url{https://nexusmutual.io/assets/docs/nmx_white_paperv2_3.pdf}.

\bibitem{insure}
inSureDAO, ``insure ecosystem: Decentralized insurance platform,'' \url{https://insuretoken.net/whitepaper.html}.

\bibitem{albrecht2017fundamental}
P.~Albrecht and M.~Huggenberger, ``The fundamental theorem of mutual insurance,'' \emph{Insurance: Mathematics and Economics}, vol.~75, pp. 180--188, 2017.

\bibitem{feng2022unified}
R.~Feng, M.~Liu, and N.~Zhang, ``A unified theory of decentralized insurance,'' \emph{Available at SSRN 4013729}, 2022.

\bibitem{riskHarbor}
M.~Resnick, R.~Ben-Har, D.~Patel, and A.~Bipin, ``Risk harbor v2,'' 2022, \url{https://github.com/Risk-Harbor/RiskHarbor-Whitepaper/blob/main/Risk%20Harbor%20Core%20V2%20Whitepaper.pdf}.

\bibitem{opium}
O.~team, ``Opium protocol whitepaper,'' 2020, \url{https://github.com/OpiumProtocol/opium-contracts/blob/master/docs/opium_whitepaper.pdf}.

\bibitem{cozy}
Cozy.Finance, ``Cozy finance developer docs,'' 2020, \url{https://docs.cozy.finance/}.

\bibitem{etherisc}
Etherisc, ``Whitepaper,'' 2022, \url{https://docs.etherisc.com/learn/whitepaper-en}.

\bibitem{unslash}
Unslashed.Finance, ``Insurance for decentralized finance,'' 2021, \url{https://documentation.unslashed.finance/}.

\bibitem{nsure}
Nsure.Network, ``Open insurance platform for open finance,'' 2020, \url{https://nsure.network/Nsure_WP_0.7.pdf}.

\bibitem{cover}
``Cover protocol,'' \url{https://github.com/CoverProtocol}.

\bibitem{enzyme}
``Enzyme finance,'' \url{https://docs.enzyme.finance/}.

\bibitem{riskHarborv1}
R.~Ben-Har, D.~Patel, A.~Su, and M.~Resnick, ``Risk harbor v1,'' \url{https://github.com/Risk-Harbor/RiskHarbor-Whitepaper/blob/8b41ca8f89ce116db725f405aeacddc18b9bd6b2/Risk%20Harbor%20V1%20Whitepaper.pdf}.

\end{thebibliography}
\end{document}